\documentclass{iopart}
\usepackage[utf8]{inputenc}
\usepackage{todonotes,color}
\usepackage{graphicx}

\newcommand{\cm}[0]{cm$^{-1}$}

\begin{document}

\title[Hydrogen Bonding in Protic Ionic Liquids]{Hydrogen Bonding in Protic Ionic Liquids: Structural Correlations, Vibrational Spectroscopy, and  Rotational Dynamics of Liquid Ethylammonium Nitrate}

\author{Tobias Zentel$^1$, Viviane Overbeck$^2$, Dirk Michalik$^2$, Oliver K\"uhn$^1$, Ralf Ludwig$^2$}

\address{$^1$Institut f\"{u}r Physik, Universit\"{a}t Rostock, Albert-Einstein-Str. 23-24, D-18059 Rostock, Germany}
\address{$^2$Abteilung f\"ur Physikalische Chemie, Institut f\"{u}r Chemie, Universit\"{a}t Rostock, Dr.-Lorenz-Weg 1, D-18059 Rostock, Germany}

\ead{oliver.kuehn@uni-rostock.de} \vspace{10pt}
\begin{indented}
\item[]July 26, 2017
\end{indented}

\begin{abstract}
The properties of the hydrogen bonds in ethylammonium nitrate are analyzed by using molecular dynamics simulations and infrared as well as nuclear magnetic resonance experiments.  Ethylammonium nitrate features a flexible three-dimensional network of hydrogen bonds with moderate strengths, which makes it distinct from related triethylammonium-based ionic liquids. First, the network's flexibility is manifested in a not very pronounced correlation of the hydrogen bond geometries, which is caused by rapid interchanges of bonding partners. The large flexibility of the network leads to a substantial broadening of the mid-IR absorption band, with the contributions due to N-H stretching motions ranging from 2800 to 3250~\cm. Finally, the different dynamics are also seen in the rotational correlation of the N-H bond vector, where a correlation time as short as 16.1~ps is observed.
\end{abstract}

%
%
\submitto{\JPB}
%
%
%

%
%

\section{Introduction}
Ethylammonium nitrate (EAN) has been synthesized more than hundred years ago by Paul Walden~\cite{walden14_405}. EAN is formed by the neutralization of ethylamine with concentrated nitric acid. It consists solely of ions, namely ethylammonium and nitrate, and melts significantly below 100$^\circ$C, i.e.\ at 13-14$^\circ$C. Thus, EAN can be regarded as the first room temperature ionic liquid (IL)~\cite{plechkova08_123,greaves08_206}. EAN belongs to the subset of protic ionic liquids (PILs)~\cite{angell07_1228,fumino09_3184,kruger10_101101,hayes12_7468}, which are of practical interest due to their ongoing development for  high-performance Li-ion cells in electric vehicles, instead of the flammable carbonate electrolytes~\cite{menne13_39}, as well as electrolytes for advanced fuel cells~\cite{markusson07_8717}. Recent studies mainly addressed the structure and thermodynamics of EAN. This archetypical PIL forms a three-dimensional hydrogen bonded (H-bonded) network, similar to that of water. In principle, these structural features were supported by molecular dynamics (MD) simulations~\cite{hayes12_7468}. In addition, symmetric and asymmetric stretching modes of the H-bonds (HBs) were observed in far infrared (FIR) spectra~\cite{fumino09_3184}.  EAN evaporates as contact ion pairs leading to low vapor pressures and an enthalpy of vaporization of about 105.3 kJmol$^{-1}$~\cite{emelyanenko14_11640}. This stability makes EAN attractive as an electrolyte for new generation fuel cells up to 419~K. In contrast, EAN evaporates as molecules near the boiling point at 513~K. Although structure and thermodynamics of EAN are widely studied, a detailed analysis of the HB network and the related dynamical information is still lacking. This is somewhat surprising because the ``flickering HB network'', similar to that of water, suggests strongly enhanced dynamics of molecular vectors that are involved in H-bonding compared to those in the one-dimensional H-bonded PIL analogues~\cite{strauch16_17788}.       

For EAN, wherein the proton exchange is fast compared to the time scale of nuclear magnetic resonance (NMR) spectroscopy, the reorientational correlation times for the N-H molecular vector, $\tau_{\rm NH}$, can  only be obtained from $^{15}$N-enhanced proton relaxation time experiments. To avoid the demanding and costly synthesis of isotopic substituted compounds, one can instead measure the deuteron relaxation times of the N-D deuterons. This relaxation mechanism is strong and purely intramolecular in nature. However, proper correlation times, $\tau_{\rm NH}$, can be obtained from the relaxation times, $T_1$, only, when reliable deuteron quadrupole coupling constants, $\chi_{\rm D}$, are known. Unfortunately, this is not the case for ILs, neither for the solid, the liquid nor the gas phase. For this reason we recently developed a method for deriving $\chi_{\rm D}$ from a relation between density functional theory (DFT) calculated $\chi_{\rm D}$ (for N-D) and proton chemical shifts $\delta^1$H (for N-H) within clusters of ILs~\cite{strauch16_17788,ludwig98_9312}. Thus, with simple measurements of the proton chemical shifts,  one can then determine accurate $\chi_{\rm D}$ values for the liquid phase.  

MD simulations  are an established tool to study ILs~\cite{Maginn09_373101,kirchner09_213,kirchner15_202}. To get a reliable description of the complex atomic motions intrinsic to HBs, methods beyond simple force field-based MD  are needed. In principle, DFT-based ab initio MD would be the method of choice, but it is hampered by the high computational costs~\cite{kirchner15_202,zahn10_124506}. An alternative is provided by the self-consistent charge DFT-based tight-binding approach (DFTB)~\cite{Elstner98_7260}. Although applications of DFTB to the dynamics of ILs are scarce, available simulations provided evidence, e.g., for reliable prediction of  X-ray diffraction data~\cite{Addicoat14_4633} and vibrational spectra as well as geometric correlations related to the HB motion of PILs~\cite{zentel16_234504,zentel17_a}. 

In the present work, we continue our investigation of the molecular dynamics of HBs in alkylammonium nitrate PILs~\cite{zentel16_234504,zentel17_a,zentel17_56} by focussing on liquid EAN. Using a combined experimental and theoretical approach, the N-H$\cdots$O HBs will be analyzed in terms of their geometric correlations,   FIR and mid infrared (MIR) signatures as well as the rotational correlation of the N-H bond  vectors. It is the purpose of this work to provide detailed information about the structure and dynamics of the very first IL EAN, with particular focus on the HB network, which resembles that of water. Section \ref{sec:methods} starts with an outline of the DFTB MD simulation protocol. Next, the method for obtaining the reorientational correlation time, $\tau_{\rm NH}$, for N-H/D bonds in EAN is sketched, which is based on the $\delta^1$H  versus $\chi_{\rm D}$ relation calculated for EAN clusters. Further, experimental details for NMR and FIR/MIR measurements are provided. In Section \ref{sec:results} results for   structure, infrared (IR) spectra, and  rotational correlation time are discussed. A summary is provided in Section \ref{sec:summary}.
\section{Methods}
\label{sec:methods}
\subsection{Molecular Dynamics}
A simulation box consisting of 60 EAN ion pairs and 900 atoms with periodic boundary conditions was built from 10 randomly positioned hexamer clusters. 
To obtain a bulk structure, the system was equilibrated  using the force field parameters from \cite{canongialopes04_16893,canongialopes04_2038}. To converge the box size a  1.5~ns simulation in the NPT ensemble at 300~K was performed using a  Parrinello-Rahman barostat. The resulting density of 1.244~g/cm$^3$ is very close to the literature value of 1.216~g/cm$^3$ \cite{greaves08_206}. To control the temperature in the subsequent NVT simulation (at 300~K), a Nos\'e-Hover thermostat with chain length 10 is used. The length of the NVT trajectory has been  5~ps. The force field-based MD simulations were performed with the Gromacs 4.5.5 software~\cite{hess08_435}.

The final equilibrated force field structure  was used as input for subsequent DFTB simulations with the DFTB+ software package~\cite{aradi07_5678}. Here, 3rd order as well as dispersion  corrections have been used together with the  3-ob  Slater-Koster parameter set~\cite{gaus11_931,yangy07_10861,gaus13_338,elstner01_5149}.  A 25 ps NVT trajectory has been simulated (time step 0.5 fs), which was used to sample starting points  for NVE trajectories every 5~ps. The NVE trajectories had a length of 30 ps with a time step of 0.5 ps.

The IR spectrum, $I(\omega)$, is obtained by Fourier transformation of  the dipole moment trajectory, $\vec{\mu}(t)$, using a standard Kaiser window function, $\kappa(t)$, with Kaiser parameter 10,  i.e. \cite{kaiser80_105}
  \begin{equation}
I(\omega) = \omega^2 \langle | \int\limits_{0}^{T} \exp(-i \omega t)   \vec{\mu}(t) \kappa(t)   \mathrm{d}t \ |^2 \rangle \, .
\label{eq:IRsignal}
\end{equation}
Here, the canonical ensemble average denoted by $\left< \cdots \right> $ is obtained by sampling initial configurations for  NVE trajectories as described above. The window function ensures that the dipole is damped to zero at the edges. The Kaiser parameter controls the time scale of the decay and is chosen large enough to avoid artificial broadening of  the spectra.
The Mulliken charges from DFTB simulations are used to calculate the dipoles of the individual ions and  of the $\mathrm{NH_3}$ and $\mathrm{CH_3}$ groups. The total box dipole is obtained as the sum of the dipoles of the individual ions.
Note that the  dipole of an ionic system is not uniquely defined, as discussed in Ref.~\cite{Brehm12_5030}. Here, we employed the center of mass of the considered subsystem as the reference for the dipole calculations.
The spectra presented below are smoothed by convolution with a Gaussian function with width parameter $\sigma= 25 \ \mathrm{cm^{-1}}$ in the MIR and $\sigma= 12.5 \ \mathrm{cm^{-1}}$ in the FIR region.

\subsection{Reorientational Correlation Time}
For the determination of the reorientational correlation time of the N-H molecular vectors, $\tau_{\rm NH}$,  we measured the deuteron relaxation time ($T_1$)$_{\rm D}$ of the deuterated species [EtND$_3$][NO$_3$] at 303~K. We have chosen quadrupolar relaxation because it is the strongest relaxation mechanism and presents a purely intramolecular process. Deuteron nuclear magnetic relaxation is driven by the interaction of the electrostatic quadrupole moment, $eQ$, of the deuteron nucleus with the main component of the electric field gradient (EFG) tensor at the nucleus, $q_{zz}$, generated by the electron distribution surrounding the nucleus along the N-H bonds. The relaxation rate ($1/T_1$)$_{\rm D}$ is given by \cite{abragam61_,farrar87_}
\begin{equation}
	\left(\frac{1}{T_1}\right)_{\rm D}=\frac{3}{10}\pi^2 \left( 1+ \frac{\eta_{\rm D}^2}{3}\right)\chi_{\rm D}^2\left( \frac{\tau_{\rm NH}}{1+\omega_0^2\tau_{\rm NH}^2} +\frac{4\tau_{\rm NH}}{1+4\omega_0^2\tau_{\rm NH}^2}\right) \, ,
\end{equation}
where $\chi_{\rm D}$ is the deuteron nuclear quadrupole coupling constant and $\eta_{\rm D}=(q_{xx}-q_{yy})/q_{zz}$ is the related asymmetry parameter. If the relaxation process does not depend on frequency and the extreme narrowing condition, $\omega_0\tau_{\rm NH} \ll 1$, holds, this equation simplifies to
\begin{equation}
	\left(\frac{1}{T_1}\right)_{\rm D}=\frac{3}{2}\pi^2 \left( 1+ \frac{\eta_{\rm D}^2}{3}\right)\chi_{\rm D}^2 \tau_{\rm NH} \, .
	\label{eq:t1}
\end{equation}
Note that we tacitly assumed that the reorientational correlation time of the principal axis of the deuterium EFG, actually entering these equations, is identical to $\tau_{\rm NH}$~\cite{huber85_4591}. Additionally, experimental data indicate that there is a negligible isotope effect for the rotational diffusion \cite{wulf07_323,wulf07_2265}.

Before the reorientational correlation times  can be determined from the quadrupolar relaxation rates, the deuteron quadrupole coupling constant  must be known. Despite what the term `coupling constant' implies, $\chi_{\rm D}$ has been shown to be temperature and solvent dependent 
\cite{ludwig94_6684,ludwig95_6941,ludwig95_329,ludwig91_89,ludwig94_313,ludwig95_19,gordalla86_817,gordalla91_975}.
Thus $\chi_{\rm D}$ is a sensitive probe for H-bonding. Unfortunately, for EAN the deuteron quadrupole coupling constant is not known at all. Thus, we used an approach for the determination of the proper deuteron quadrupole coupling constant in the liquid phase, which has been successfully applied for N-D coupling constants in other PILs   and is briefly described here \cite{strauch16_17788}.

The relation between the principal component of the EFG tensor, $eq_{zz}$, and the deuteron nuclear quadrupole coupling constant, $\chi_{\rm D}$, is given by ($h$ is Planck's constant)
\begin{equation}
	\chi_{\rm D}=\left(\frac{eQeq_{zz}}{h}\right) \, .
\end{equation}
In principle, the deuteron quadrupole coupling constant  can be now obtained by multiplying calculated EFG  by a calibrated nuclear quadrupole moment, $eQ$. The latter  is obtained by plotting the measured gas phase quadrupole coupling constants from microwave spectroscopy versus the calculated EFGs for small molecules such as CD$_4$, CD$_3$OH, ND$_3$ etc. as described by Huber et al.
\cite{huber85_4591,eggenberger92_5898,searles_}.
 The slope gives a reasonable value of $eQ =295.5$~fm$^2$. This approach holds regardless of whether gas phase molecules, H-bonded clusters or IL complexes are investigated. 

Recently, we could show for a large set of DFT-calculated clusters of trialkylammonium-based PILs that there exists a linear correlation between the calculated proton chemical shifts  and the calculated  deuteron quadrupole coupling constants~\cite{strauch16_17788}. The advantage now is that $\delta^1$H can be easily measured in the liquid phase, and thus, by virtue of the computationally established linear correlation, provides access to  the  $\chi_{\rm D}$ values. Thus, we applied this approach to the present EAN. First, we obtained the proton chemical shifts, $\delta^1$H, and the quadrupole coupling constants, $\chi_{\rm D}$, from DFT calculated clusters including $n=2, 4, 6, 8$ ion-pairs. 
Throughout, the geometries of the clusters as well as the proton chemical shifts and the EFGs  of the deuterons were calculated using Gaussian 09~\cite{frisch09} at the B3LYP-D3/6-31+G* level of theory, including D3 dispersion correction as introduced by Grimme et al.~\cite{grimme10_154104,ehrlich11_3414,grimme16_5105}.
 The proton chemical shifts were referenced against  TMS (tetramethylsilane), as it was done in the experiment (see below). The $\chi_{\rm D}$ values were derived by multiplying the calculated main components of the EFG tensor $q_{zz}$ with the calibrated nuclear quadrupole moment $eQ =295.5$~fm$^2$. 
The asymmetry parameter of the EFG  for N-D, $\eta_{\rm D}$, was found to be negligible and did not have to be considered in (\ref{eq:t1}). This has been shown for molecular systems such as ammonia, formamide, N-methyl formamide and N-methyl acetamide 
\cite{ludwig95_5118,ludwig95_3636,ludwig97_8861,ludwig97_499,ludwig98_197,ludwig98_205}.
 
 The reorientational correlation times, $\tau_{\rm NH}$, thus obtained from (\ref{eq:t1}) will be compared with   simulation results. To this end, the rotational time correlation function, $C(t)$, is computed according to
\begin{equation}
C(t) = \left< P_2( \vec{u}(0) \vec{u}(t)) \right>  \, .
\label{eq:rotTCF}
\end{equation}
Here, $P_2$ is the second Legendre polynomial and ${\vec u}(t)$ the unit vector along the N-H bond~\cite{lawrence03_264}. 
From $C(t)$ the rotational correlation time  follows as
\begin{equation}
\tau_{\rm NH} = \int_0^{\infty} \left< P_2( \vec{u}(0) \vec{u}(t) \right> \mathrm{d}t \, .
\label{eq:tau}
\end{equation}

 %
\subsection{Experimental Methods}
The  EAN samples were of commercial origin (IOLITEC) with purity degrees of  $>97$\%. Prior to experiments, the sample was subjected to vacuum evaporation at 333 K for more than 72 hours to remove possible traces of solvents and moisture. Once purified, the sample was transferred under inert argon atmosphere and stored in a hermetically sealed bottle. The purity of the EAN sample was confirmed with $^1$H and $^{13}$C NMR (see Supplement). The residual water concentration in the sample for combustion calorimetry was determined by Karl Fischer titration and was lower than 100 ppm. The $^1$H NMR spectra of the protonated EAN were recorded on a Bruker Avance 500 MHz spectrometer using 5 mm probes. The $\delta^1$H chemical shifts were measured versus TMS.  The EAN sample was deuterated by H/D exchange in D$_2$O and dried again. Longitudinal magnetic relaxation times $T_1$ were measured with the same spectrometer at a resonance frequency of $\nu_0=\omega_0/2\pi$= 76.7~MHz employing the inversion recovery ($180^\circ - \tau - 90^\circ$) pulse sequence~\cite{farrar87_} (see Supplement). $T_1$ is estimated to be accurate to within $\pm$2\%. Temperature calibrations were carried out using methanol and  ethylene glycol NMR thermometers.	

The  FIR measurements were performed with a Bruker Vertex 70 FTIR spectrometer. The instrument was equipped with an extension for measurements in this spectral region. This equipment consists of a multilayer mylar beam splitter, a room temperature DLATGS detector with preamplifier and polyethylene   windows for the internal optical path. The accessible spectral region for this configuration lies between 30 and 680~\cm. The spectrometer was purged continuously with dry air during the experiments in order to minimize contributions from atmospheric water vapor. Further reduction of these signals was achieved by utilization of telescope tubes with polyethylene windows reducing the optical path in the sample chamber to a minimum. In this way, fluctuations in the atmospheric water content can be prevented. The L.O.T.-Oriel cell was equipped with polyethylene windows. The 0.012 mm path length was realized by tin spacers. For each spectrum 100 scans at a spectral resolution of 1~\cm{} were recorded. The experimental spectrum shown in Figure \ref{fig:EANfarIR} has been published elsewhere~\cite{fumino09_3184}. 
The MIR measurements were performed with a Bruker Vector 22 FTIR spectrometer. An L.O.T.-Oriel variable-temperature cell equipped with CaF$_2$ windows was used. For each spectrum 128 scans were recorded at a spectral resolution of 1~\cm.
\section{Results and Discussion}
\label{sec:results}
\subsection{H-Bond Geometries}
%
%

EAN facilitates multiple HB opportunities and thus is capable of forming a complex network structure, which has been discussed to be comparable to that of water \cite{fumino09_3184,evans81_481}. There exist three acceptor oxygen atoms per anion and three donor positions per cation, so that in principle a fully saturated HB network is possible. Note that the trajectory simulations discussed below did not exhibit any proton transfer events leading to a stable product of two neutral molecules. This might be related to the relatively short propagation time.

\begin{figure}
\centering
\includegraphics[scale=0.5]{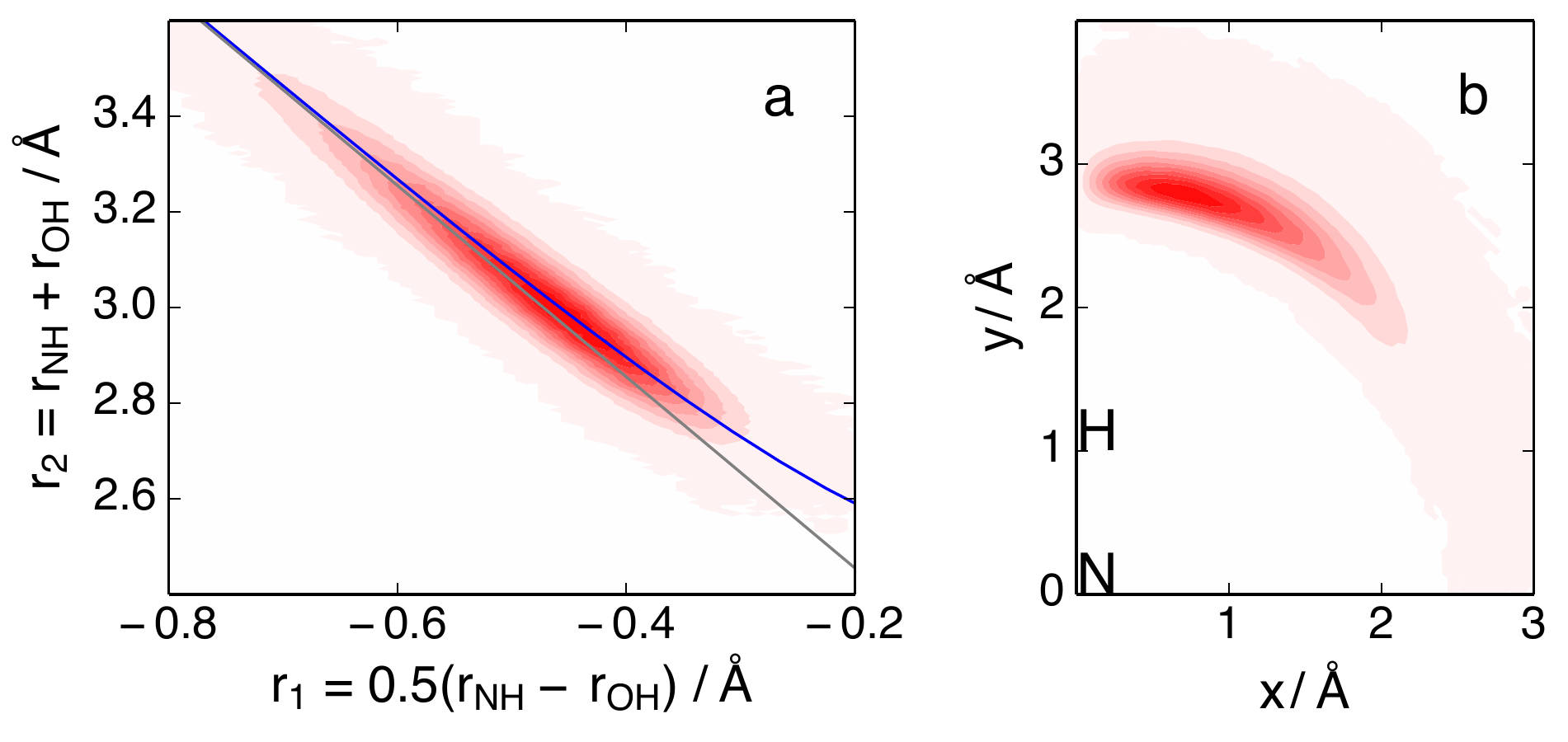}
\caption{(a) HB geometric correlations of EAN, expressed by using internal HB coordinates as indicated. (b) Distribution of angles HB along the NVT trajectory. Shown is the accepting oxygen's position in the plane defined by the positions of the N, H and O atom. The nitrogen is positioned at the origin and the approximate hydrogen position is marked by 'H'.}
\label{fig:Limbach}
\end{figure}

%
Nitrate is a strongly interacting anion and we assume all HB donors to be involved in H-bonding. In the following a HB is counted always between an ammonium hydrogen and its closest oxygen.  The average HB distances are $r_{\mathrm{NO}}$=2.91~$\mathrm{\AA}$, $r_{\mathrm{NH}}$=1.04~$\mathrm{\AA}$, and    $r_{\mathrm{OH}}$=2.02~$\mathrm{\AA}$. These values are in the range  expected  for moderate strong HBs \cite{Hunt15_1257}. Further, EAN exhibits a bent HB geometry with an average angle, $\alpha$, between the NH and OH vector of 34$^{\circ}$ (see also figure \ref{fig:Limbach}). Only 16\% of the HBs are linear ($\alpha <15^\circ$). Hayes et al. investigated the structure of various PILs including EAN with Empirical Potential Structure Refinement (EPSR) and Monte Carlo methods \cite{Hayes13_4623}. They found 12\% of the HBs being linear ($\alpha <15^\circ$) and 1.6  oxygen atoms per ammonium hydrogen. In total this classifies the  HBs in EAN as being substantially bifurcated. 
 
The continuous lifetime of a H-bonded ion pair is  \mbox{$\tau^{\mathrm{HB}}_{\mathrm{cont}}$=0.14~ps} (average lifetime of non-interrupted HBs) and the intermittent lifetime is \mbox{$\tau^{\mathrm{HB}}_{\mathrm{inter}}$=8.6~ps} (total average lifetime along trajectories). It is not unusual that the two lifetimes do not match \cite{lawrence03_264}. However, it should be noted that in the current analysis only one HB per ammonium hydrogen is possible. Thus, thermal motion of a bifurcated HB might, according to the distance criterion, be counted as bond breaking and reforming. In this case, the HB lifetimes would be  underestimated.

\begin{figure}
\centering
\includegraphics[width=0.7\textwidth]{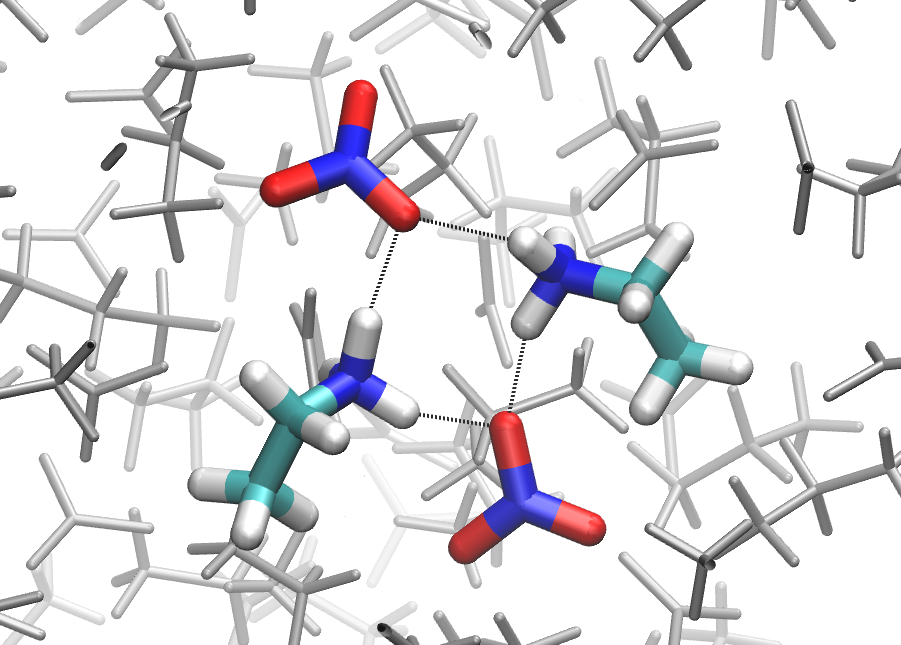}
\caption{HB ring-like structures such as the one shown here are an ubiquitous structural feature found for the  EAN equilibrium trajectory.} 
\label{fig:EANring}
\end{figure}

The HBs of one ammonium group are almost exclusively formed with three different anions (less than 1\% of the HBs of one ammonium group are to the same counter anion). In 27\% of the cases an oxygen atom is acceptor of two HBs, leaving some of the oxygen atoms unpaired. Furthermore, there exist a number of ring-like structures, where two cations are  H-bonded to the same two anions, an exemplary structure is shown in figure~\ref{fig:EANring}. This can be either via different oxygen atoms of the same anion or via a single oxygen involved in two HBs (as shown). In the present bulk MD simulations containing 60 ion pairs, on average 18.4 rings exist, consisting of 73.6 ions. Since there are three HB possibilities per ion  it is possible for ions to be involved in more than one ring. Especially interesting are 'double rings' that form a HB network cluster consisting of two cations that share all three anions. On average 2.6 cluster structures are found. 
We can conclude that the  ring-like geometries appear to be  an abundant structural motif in EAN and can be taken as evidence for the generally assumed three-dimensional network character of the HBs in EAN.

In order to quantify geometric correlations of the HBs in EAN, the valence bond model of Pauling is used in the following~\cite{pauling47_542,Limbach04_5195,yan10_15695,yan11_1}. Here, starting from the HB distances, $r_{\mathrm{NH}}$ and $r_{\mathrm{OH}}$, bond orders are defined as $p_i = \exp(-(r_i - r_{i}^{\mathrm{eq}} )/b_i)$ with $i = \mathrm{ \{ NH, OH\}}$ and $r_{i}^{\mathrm{eq}}$ being the gas phase equilibrium bond distances. Under the constraint that the sum of the two bond orders must equal one, the two coordinates depend on each other and the HB geometry change along a proton transfer reaction can be described by a single coordinate \cite{Limbach04_5195}. This path is commonly plotted into a graph showing the correlation between the coordinates $r_1=0.5(r_{\rm NH}-r_{\rm OH})$ and $r_2=r_{\rm NH}+r_{\rm OH}$.  For the case of EAN the gas phase equilibrium distances $r_i^{\mathrm{eq}}$ are obtained from single molecule gas phase DFTB optimizations. The NH bond length is set to the average of the NH bond lengths of the optimized structure. 
\begin{table}
\centering
\begin{tabular}{|l|c|c|c|c|c|c|}
\hline & $r_{\mathrm{NH}}^{\mathrm{eq}}   $ & $ r_{\mathrm{OH}}^{\mathrm{eq}} $ & $b_{\mathrm{OH}}$ & $b_{\mathrm{NH}}$ & $\left< r_1 \right>  $ & $\left< r_2 \ \right> $ \\
\hline 
EAN  &1.03  &0.98 & 0.33 & 0.30  & -0.49 &3.06\\
\hline
TEAN &  1.04 &0.98 & 0.35 & 0.32&-0.47 &3.06\\
\hline
\end{tabular}
\caption{Parameters of the Pauling valence bond order model as obtained  for EAN and TEAN~\cite{zentel16_234504} (all values in \AA).}
\label{tab:Limbach}
\end{table}
The bond order decay parameters were changed simultaneously starting from the values reported for HBs in crystal structures \cite{steiner98_7041} until a good visual fit with the DFTB data sampled along the trajectory was observed. This results in $b_{\mathrm{OH}}=0.3$~\AA{} and $b_{\mathrm{NH}}=0.33$~\AA.  All parameters of the Pauling model are collected in table~\ref{tab:Limbach}.  In passing we note that the same values were found for EAN gas phase cluster simulations~\cite{zentel17_a}. 

The resulting  reaction path is drawn in figure~\ref{fig:Limbach} (blue line) together with the distribution of geometries according to the trajectory. If the distances $r_{\rm NH}$ and $r_{\rm  OH}$ were fully uncorrelated, the geometries would lie on a straight line with  slope 2, plotted for reference as a gray line. From the apparent deviation from linearity one can conclude on a weak correlation between the bond distances across the HB.

This analysis was previously performed for triethylammonium nitrate (TEAN) using a similar setup~\cite{zentel16_234504}. The resulting parameters of the valence bond order model are also given in table~\ref{tab:Limbach}. In particular the  decay parameters were found to be $b_{\mathrm{NH}} =0.35~\mathrm{\AA}$ and $b_{\mathrm{OH}}=0.32~\mathrm{\AA}$. Thus, judging from the geometric correlations the HBs in EAN are slightly weaker than in TEAN. This can be attributed to the HB angle. Compared with TEAN, in EAN not only the average angle is larger (26$^\circ$ versus 34$^\circ$), but also the distribution is broader (cf. figure~\ref{fig:Limbach}b).
\subsection{IR Spectra}
\begin{figure}
\centering
\includegraphics[scale=0.5]{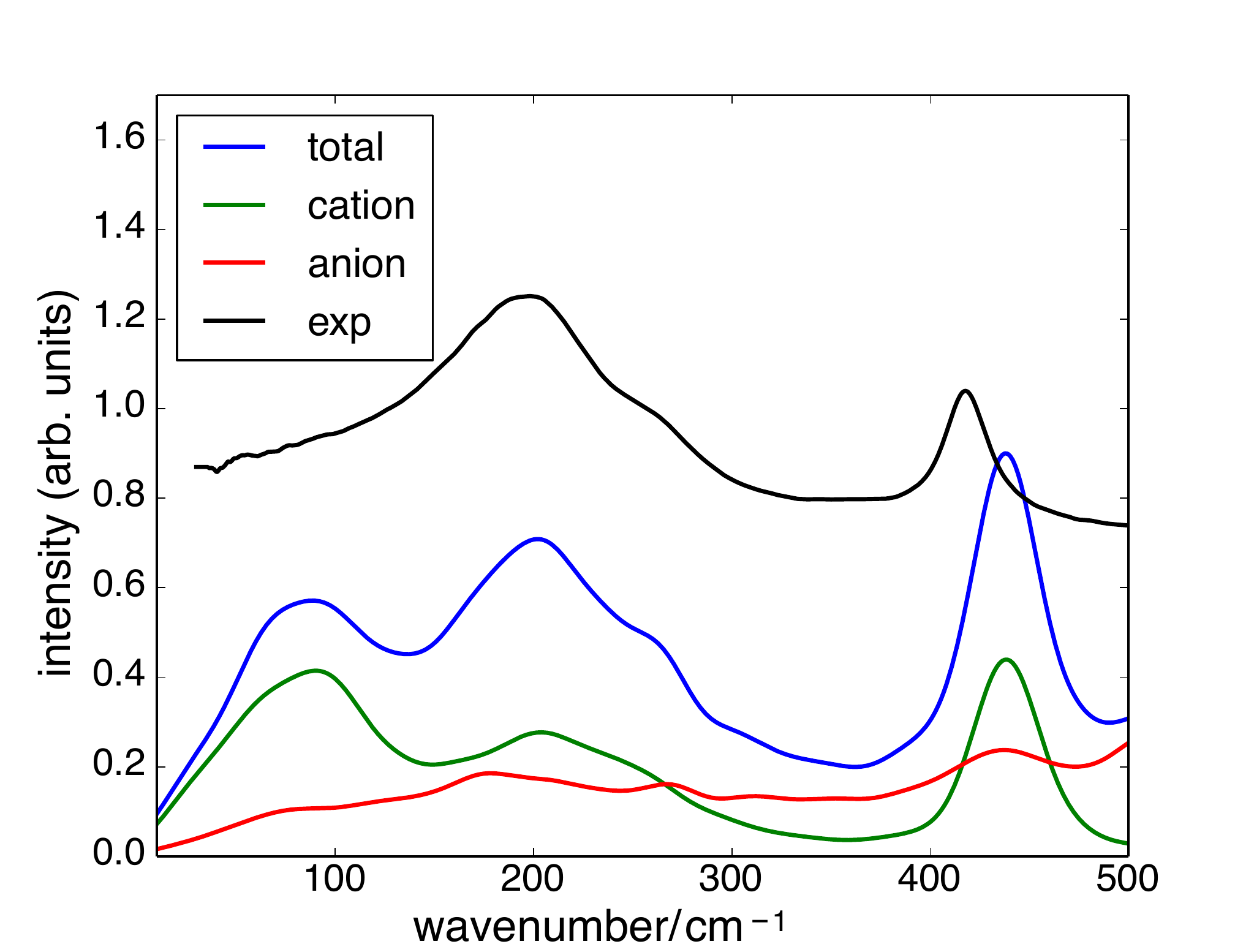}
\caption{FIR spectrum of EAN according to the MD simulation (blue) and experiment (black,  offset for better visibility). Also shown are spectra calculated separately for the anion (red) and cation (green) species.}
\label{fig:EANfarIR}
\end{figure}

%
In figure~\ref{fig:EANfarIR} the FIR spectrum calculated via (\ref{eq:IRsignal}) is presented together with an experimental spectrum taken from  \cite{fumino09_3184}. The calculated spectrum (blue) shows three peaks in the FIR region, i.e. at 92~$\mathrm{cm^{-1}}$, 201~$\mathrm{cm^{-1}}$ (with a shoulder at 260~$\mathrm{cm^{-1}}$), and 439~$\mathrm{cm^{-1}}$. To understand the origin of these peaks, the spectrum has been additionally computed from the dipoles of the anions, ${\vec \mu}_{\mathrm{anion}}$ (red), and cations, ${\vec \mu}_{\mathrm{cation}}$ (green), only. In general the intensities of the cation and anion spectra  do not add up to the full spectrum, because cross-terms ${\vec \mu}_{\mathrm{cation}}\cdot{\vec \mu}_{\mathrm{anion}}$ originating from the square of the total dipole,  ${\vec \mu}_{\mathrm{cation}} + {\vec \mu}_{\mathrm{anion}}$, in (\ref{eq:IRsignal}) are not included in the separate spectra. 
The peak at the lowest frequency  stems almost exclusively from cationic motion and is attributed to be essentially due to dispersion interactions, which usually are dominated by alkyl tails, with smaller contributions due to interionic HB motion.  Note that this peak was not present in previous hexamer gas phase cluster simulations, since in this case the alkyl chains are pointing outwards what reduces their interaction~\cite{zentel17_a}. 

The peak at 201 $\mathrm{cm^{-1}}$ has a low intensity in cationic as well as anionic spectra. Hence, it can be attributed to cross-terms, which suggests that the band is due to a correlated cation-anion motion likely originating from interionic HBs.
The peak around 440 $\mathrm{cm^{-1}}$ was previously assigned by normal mode calculations to a N-C-C bending motion of the cation \cite{zentel17_a}. 

These assignments are in line with previous FIR experimental investigations of EAN~\cite{fumino09_3184,fumino12_6236}.
Overall the simulation results reproduce the experimental spectra very well. The largest difference is the absence of a clear  low-frequency dispersion band around 92~$\mathrm{cm^{-1}}$ in the experiment. This could be due to a larger broadening of this band in the experimental data. In fact related bands around 100~\cm{} are discernible in less flexible TEA-like systems~\cite{fumino12_6236}.

%

\begin{figure}
\centering
\includegraphics[scale=0.5]{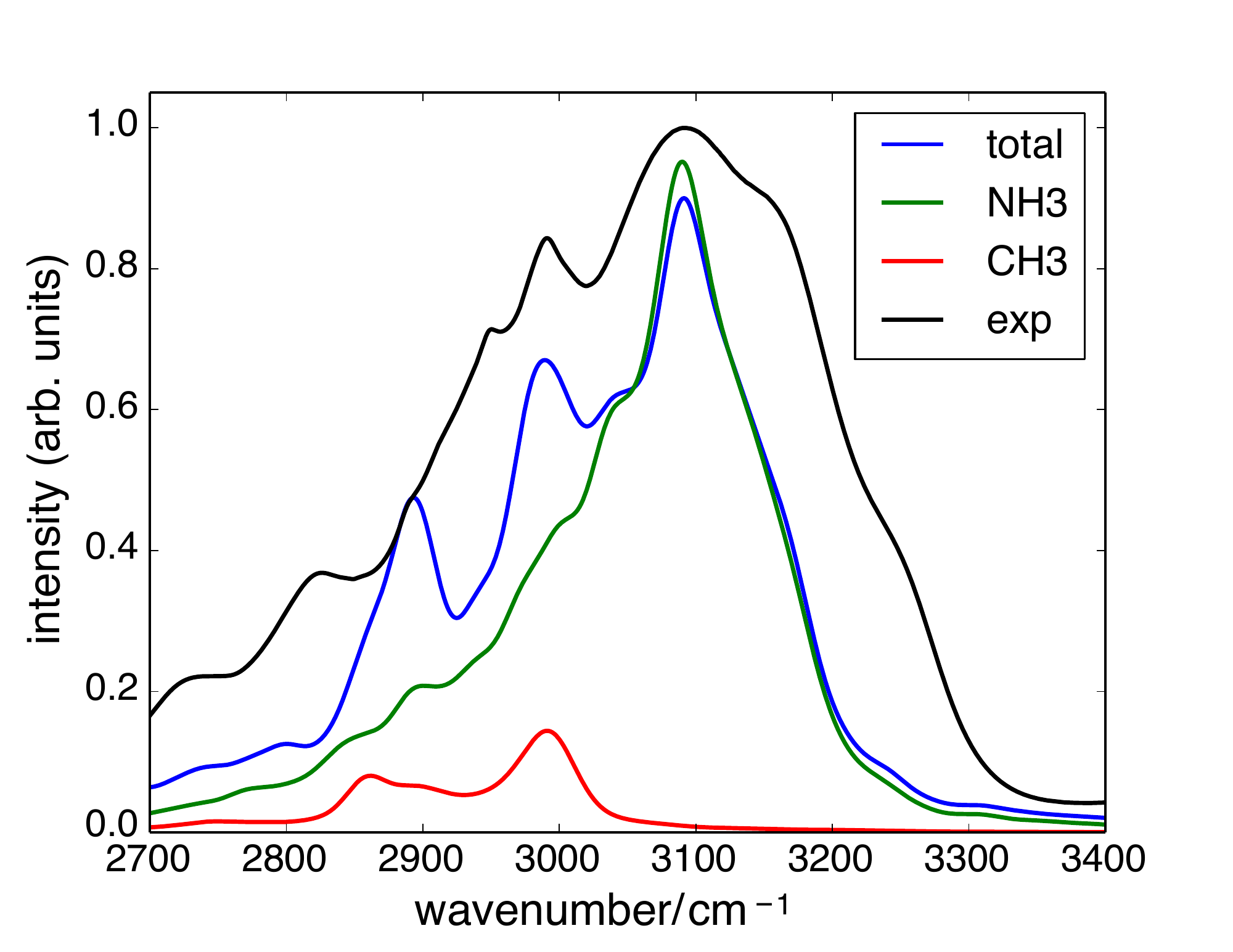}
\caption{MIR spectrum of EAN according to the MD simulation (blue) and experiment (black). Also shown are spectra calculated separately for the $\mathrm{NH_3}$ (green) and $\mathrm{CH_3}$ (red) groups.} 
\label{fig:EANIR3000}
\end{figure}

The MIR spectral region is shown with spectra from experiment (black) and simulation (blue) in figure~\ref{fig:EANIR3000}. The experimental signal covers a wide range from 2700 to 3300 $\mathrm{cm^{-1}}$ and shows a left-skewed distribution. The maximum intensity is located at 3092 $\mathrm{cm^{-1}}$ with a side peak at 3159 $\mathrm{cm^{-1}}$ and a shoulder at 3250 $\mathrm{cm^{-1}}$. Further, there are several local maxima on the low-frequency side. In general, the MIR spectrum results from the extended three-dimensional HB network, including possible  proton transfers between anions and cations.
 Due to H-bonding the symmetric and anti-symmetric NH stretching vibrations are strongly red-shifted and overlap with the corresponding C-H vibrational bands of the ethyl group. To assign the  observed structure, the separate contributions to the calculated spectrum of the $\mathrm{NH_3}$ (green) and $\mathrm{CH_3}$ (red) dipoles are also shown in figure~\ref{fig:EANIR3000}. For better visual comparison the maximum of the $\mathrm{CH_3}$-only simulation is used as reference to scale all theoretical spectra by aligning it to the position found for the respective modes in experimental Raman spectra \cite{Bodo13_144309}. This results in a scaling of the frequencies by a factor of 0.978.

Overall, the agreement between simulation and experiment is very good, i.e. position and general shape of the spectra match very well. As expected the contribution of  the polar $\mathrm{NH_3}$ groups has a much higher intensity than those of the non-polar $\mathrm{CH_3}$  groups.  The IR signal above 3050 $\mathrm{cm^{-1}}$ stems exclusively from $\mathrm{NH_3}$ motion and the maximum coincides with that of the full spectrum. Note the perfect match of the global maximum at 3092 (unscaled 3162) $\mathrm{cm^{-1}}$, mainly originating from the $\mathrm{NH_3}$ dipole, which is achieved even though only the $\mathrm{CH_3}$ peak was used to obtain the scaling factor. Additionally, the contribution from the $\mathrm{NH_3}$ dipoles shows the same left-skewed behavior as the full spectrum  and it is very broad, i.e. spanning the range  from 2800 to 3250 $\mathrm{cm^{-1}}$. The latter indicates a substantial distribution of HB strengths. The alkyl CH stretching vibrations have a much weaker intensity and span a narrower range in between two maxima at 2923 (3060) $\mathrm{cm^{-1}}$ and 2863 (2927) $\mathrm{cm^{-1}}$.


We note that the spectrum according to hexamer gas phase simulations using the DFTB method~\cite{zentel17_a} deviates from that of the present bulk case. Especially, the cluster shows a larger HB strength, as indicated by a larger blue shift in FIR and a larger  red shift in MIR of related spectral features. The results of the present simulations can be further compared with bulk phase calculation for TEAN reported in \cite{zentel16_234504}. One finds that for EAN the red shift of the NH-stretching vibrations in the MIR region is smaller, but the blue shift in the FIR region is more pronounced. At first sight, this is a surprising result since the magnitudes of both shifts are associated with the HB strength. According to the above analysis of HB geometries, EAN should feature a weaker HB than TEAN. Thus, the comparison of shifts in the FIR region is misleading insofar as the interionic motion in EAN is influenced by three possible HBs per ion, whereas there is only one for TEAN. On average this may cause a tighter potential and hence a larger blue-shift.
%
%

%

\begin{figure}
\centering
\includegraphics[scale=0.45]{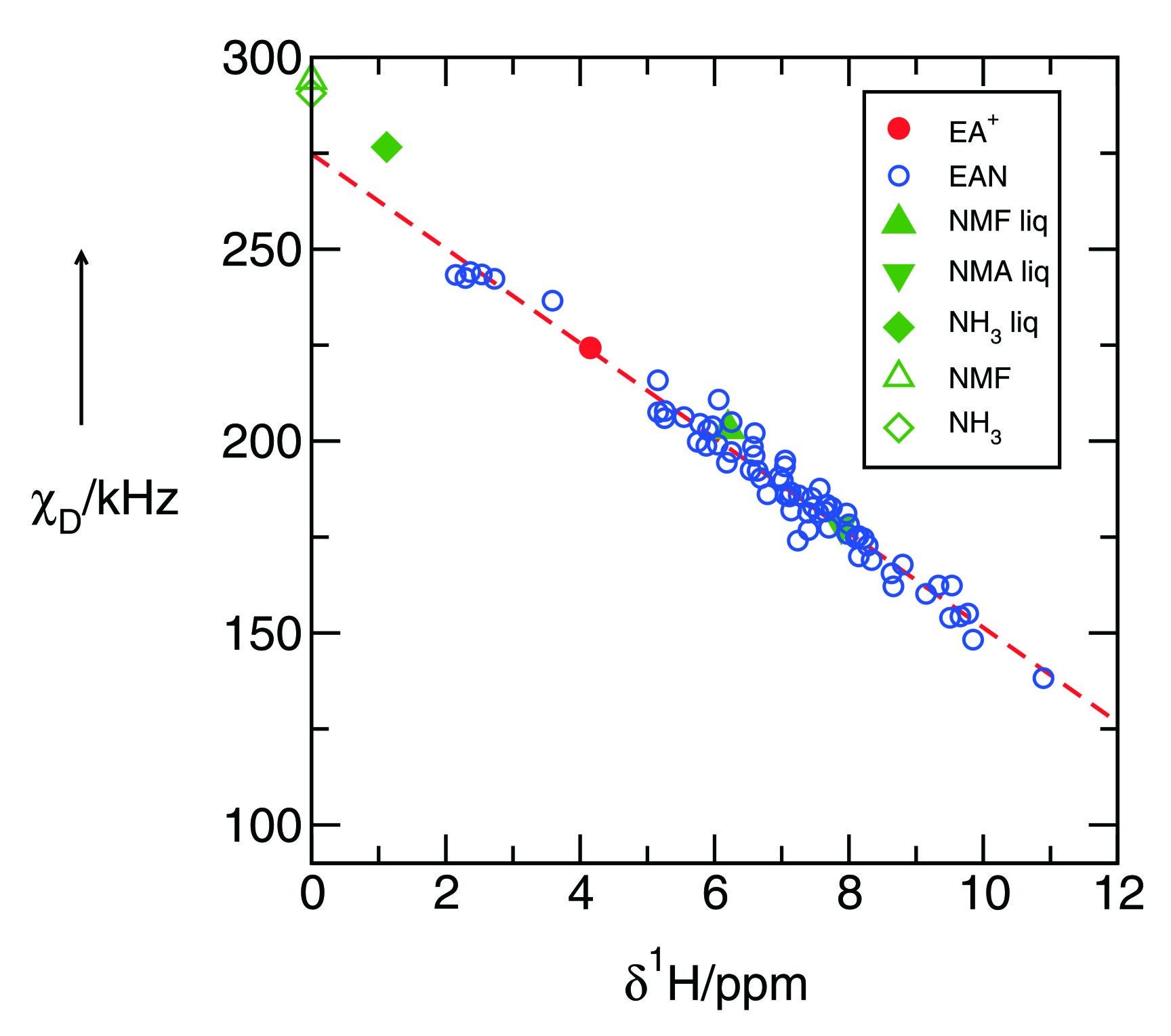}
\caption{DFT (B3LYP-D3/6-31+G*) calculated deuteron quadrupole coupling constants, $\chi_{\rm D}$, plotted versus calculated proton chemical shifts, $\delta^1$H, of the N-H group in different sized clusters EAN including $n=$2, 4, 6, and 8 ion pairs.  For proving the validity of the relation between both properties we added measured gas phase or calculated monomer values for   N-methylformamide (NMF)~\cite{ludwig97_499} and ammonia (NH$_3$)~\cite{ludwig98_197,ludwig98_205}, as well as the liquid phase values for NMF~\cite{ludwig97_499,seipelt97_1501}, NH$_3$~\cite{ludwig98_197,ludwig98_205} and N-methylacetamide (NMA)~\cite{ludwig98_9312,ludwig97_8861,seipelt97_1501}. The $\chi_{\rm D}$ values for the molecular liquids show the same linear dependence as the PILs.
} 
\label{fig:d1H-chiD}
\end{figure}

\subsection{Rotational Correlation}
In figure \ref{fig:d1H-chiD} we show the  calculated deuteron quadrupole coupling constants versus calculated proton chemical shifts for HBs taken from different sized EAN clusters. Apparently, there is  an almost linear dependence. From linear regression we obtain the relation $\chi_{\rm D}$=274.86~kHz - 12.342~$\delta^1$H kHz/ppm. 
We observe that the proton chemical shifts vary between 2 and 11~ppm depending on the strength of the NH$\cdots$O cation-anion interaction at the different  positions within the clusters. This range for the chemical shifts corresponds to deuteron quadrupole coupling constants varying from 250 down to 140 kHz.   The weaker HBs are characterized by smaller $\delta^1$H  shifts, and larger $\chi_{\rm D}$ values, and vice versa. It does not matter, whether these pairs of properties are calculated for N-H/N-D in varying sized clusters or for different configurations within these clusters, they all show a linear dependence. This is not at all surprising, because both properties, $\delta^1$H  as well as $\chi_{\rm D}$, are sensitive to local and directional interactions, such as H-bonding. 

The ultimate test for the reliability of the linear relationship between both properties in PILs is carried out by interpolating the chemical shifts to 0 ppm, indicating the absence of intermolecular interactions.  For this case we expect the calculated $\chi_{\rm D}$ values to be similar to those measured for the gas phase of similar molecules. And indeed, the estimated value of 275 kHz for $\chi_{\rm D}$ in EAN is only slightly lower than the measured gas phase and calculated monomer $\chi_{\rm D}$ values for ammonia (290.6 kHz), formamide (292 kHz) and N-methylformamide (294 kHz), respectively
 \cite{ludwig95_5118,ludwig95_3636,ludwig97_8861,ludwig97_499,seipelt97_1501}.

\begin{figure}
\centering
\includegraphics[scale=0.45]{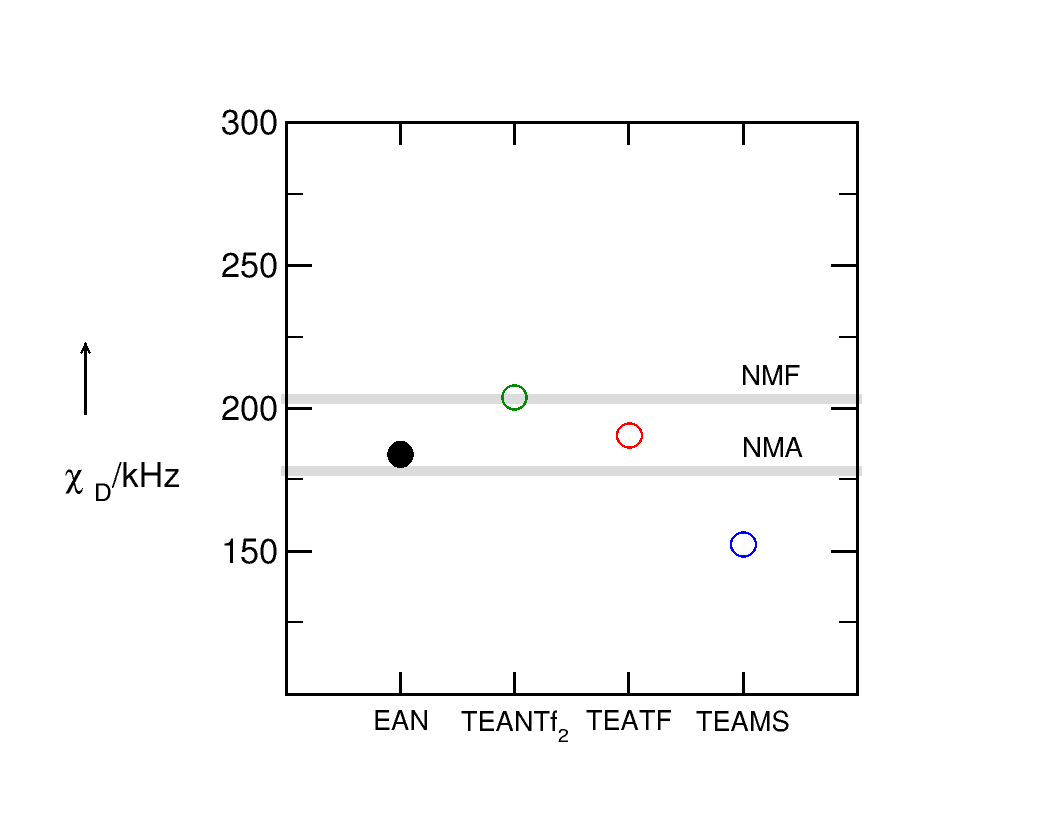}
\caption{Deuteron quadrupole coupling constant $\chi_{\rm D}$ for the N-D deuterons of EAN at 303 K. For comparison we show the $\chi_{\rm D}$ values for triethylammonium-based PILs including bis(trifluoromethylsulfonyl)-imide [NTf$_2$] (I), trifluoromethyl sulfonate [CF$_3$SO$_3$] (II), and methylsulfonate [CH$_3$SO$_3$] (III) as anions, respectively (data taken from \cite{strauch16_17788}). The grey bars indicate the $\chi_{\rm D}$ values for molecular liquids such as  NMF  and  NMA \cite{seipelt97_1501,ludwig95_5118,ludwig95_3636,ludwig97_8861,ludwig97_499,ludwig98_197,ludwig98_205}.
} 
\label{fig:XD}
\end{figure}

We now use this relation for deriving the deuteron quadrupole coupling constant of EAN in the liquid phase. From the measured proton chemical shift $\delta^1$H=7.375~ppm at 303~K we obtain a deuteron quadrupole constant of $\chi_{\rm D}=184$~kHz (see table \ref{tab:relax}). In figure \ref{fig:XD} we compare the $\chi_{\rm D}$ value for EAN with those of triethylammonium-based PILs, including bis(trifluoromethylsulfonyl)-imide [NTf$_2$], trifluoromethyl sulfonate [CF$_3$SO$_3$], and methylsulfonate [CH$_3$SO$_3$] as anions (PILs I-III). At a first glance, it seems to be surprising that EAN, which includes the strongest interacting nitrate anion, does not give the lowest $\chi_{\rm D}$ value. 
However, in the TEA-based PILs we have single, one-dimensional HBs between cation and anion, whereas in EAN a three-dimensional HB network is formed. Thus the charge transfer from the anions to N-H bonds of the cations is distributed on three bonds rather than one as it is the case for the other PILs.    Clearly, the quadrupole coupling constant is a sensitive probe for H-bonding in ILs, as it has been previously shown for molecular liquids \cite{wendt98_1077,wendt99_753,wulf06_266}.

\begin{figure}
\centering
\includegraphics[scale=0.4]{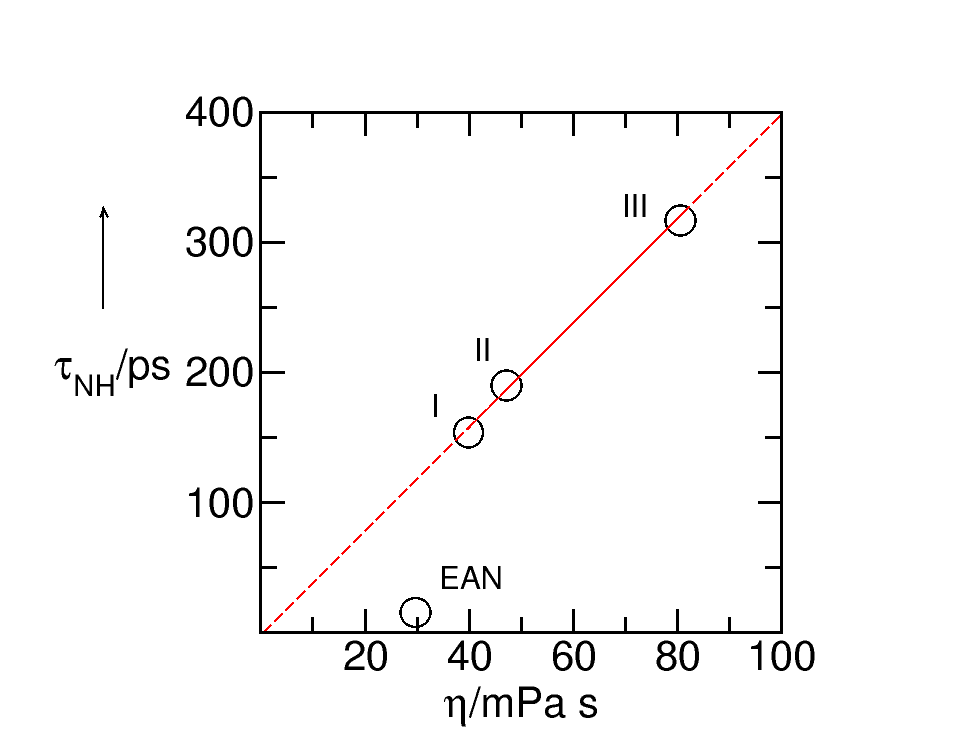}
\caption{NMR reorientational correlation times $\tau_{\rm NH}$ plotted versus viscosities $\eta$. According to the  SED  relation, the linear behavior indicates that the effective volume $V_{\rm eff}$ is similar for PILs I-III. The calculated size for the solute is that of the triethyl ammonium cation. Obviously, the NMR $\tau_{\rm NH}$ of EAN does not describe the simple hydrodynamic basis of the SED relation.
} 
\label{fig:eta_vs_tau}
\end{figure}

The obtained $\chi_{\rm D}$ value of 184 kHz can now be used for the determination of the reorientational correlation time  using (\ref{eq:t1}), which gives $\tau_{\rm NH}= 16.1$~ps at 303 K. Compared to the other PILs, which are described by single N-H bonds and weaker interacting anions, the correlation time of EAN are surprisingly low. The $\tau_{\rm NH}$ values for PILs I-III are 154.2 ps, 190.2 ps and 316.9 ps, respectively and thus one order of magnitude larger than those of EAN (see table \ref{tab:relax}). In PILs I-III the molecular motion of the cations was slowed down in the order [NTf$_2$] $<$ [CF$_3$SO$_3$] $<$ [CH$_3$SO$_3$]. For the strongly interacting [NO$_3$] anion even larger correlation times should be expected. The strongly enhanced dynamics in EAN can ony be explained by the high flexibility of the HB network and fast proton exchange. That becomes obvious if we compare the correlation times for all four PILs with the corresponding viscosities.

We recently discussed the relation between reorientational correlation times   and viscosities, $\eta$, for PILs I-III. The observed linear behavior is expected from the Stokes-Einstein-Debye (SED) relation~\cite{einstein56_,debye29_}, $\tau_{\rm NH}= \eta V_{\rm eff}/k_{\rm B}T$ if the volume/size of the solute is similar for all PILs. Here, $V_{\rm eff}$ is the effective volume obtained by multiplying the volume $V$ with the so-called Gierer-Wirtz factor~\cite{gierer53_532}, which for neat liquids has a value of about 0.16. 
 From the slope of the plot in figure \ref{fig:eta_vs_tau}, we can estimate the effective volume, $V_{\rm eff}$, to be about 0.1217~nm$^3$. If we assume a spherical shape of the solute particles, we obtain an effective radius, $R_{\rm eff}$, of about 3.07~\AA. This radius is in reasonable agreement with the one of the TEA cation, which has been calculated to be 3.22~\AA~\cite{strauch16_17788}. Since we use reorientational correlation times of the cation only, which are identical in all three PILs, a linear behavior between $\tau_{\rm NH}$  and  $\eta$ follows. That the straight line goes through zero further supports the idea that the microscopic correlation times and macroscopic viscosities describe the dynamical behavior of similar molecular species.  If we add $\tau_{\rm NH}$ and $\eta$ for EAN to figure \ref{fig:eta_vs_tau}, we observe that this relation does not hold for the smaller ethylammonium cation.      Even if we take into account, that the volume of ethylammonium is half as large as that of the TEA cation, the measured correlation time is too small to be described by the SED relation.  Obviously, the correlation time reflects the shorter life times of the HBs in the three-dimensional network of EAN.  In contrast, in PILs I-III $\tau_{\rm NH}$ reflects the rotational dynamics of the strong HB within the ion pair and is thus more suitable to describe the simple hydrodynamic basis of the SED relation.

\begin{table}
\centering
\begin{tabular}{|l|c|c|c|c|c|c|}
\hline & $T_1$ /ms & $T_1^{-1}$ /s$^{-1}$ & $\delta^1$H /ppm & $\chi_{\rm D}$ /kHz & $\tau_{\rm NH}$ /ps & $\eta$ /mPas \\
\hline 
EAN  &		124  &	8.03& 	7.38 & 	183.8  & 	16.1 &	29.65 \cite{zarrougui15_686} \\
TEANf$_2$ &	11 &	94.88 &		5.96 &	203.9 &		154.2&	39.78 \\
TEATf &		10 & 	95.33 & 	7.22 & 	184.0 	& 	190.2 & 47.20\\
TEAMs & 	9&	116.81&		8.88 &	157.8 &		316.9 & 	80.63\\
\hline
\end{tabular}
\caption{Relaxation times $T_1$,  relaxation rates $1/T_1$, NMR chemical shifts $\delta^1$H, deuteron quadrupole coupling constants $\chi_{\rm D}$, and reorientational correlation times $\tau_{\rm NH}$ of the N-H/D bonds in EAN and in PILs I-III. }
\label{tab:relax}
\end{table}

\begin{figure}
\centering
\includegraphics[scale=0.5]{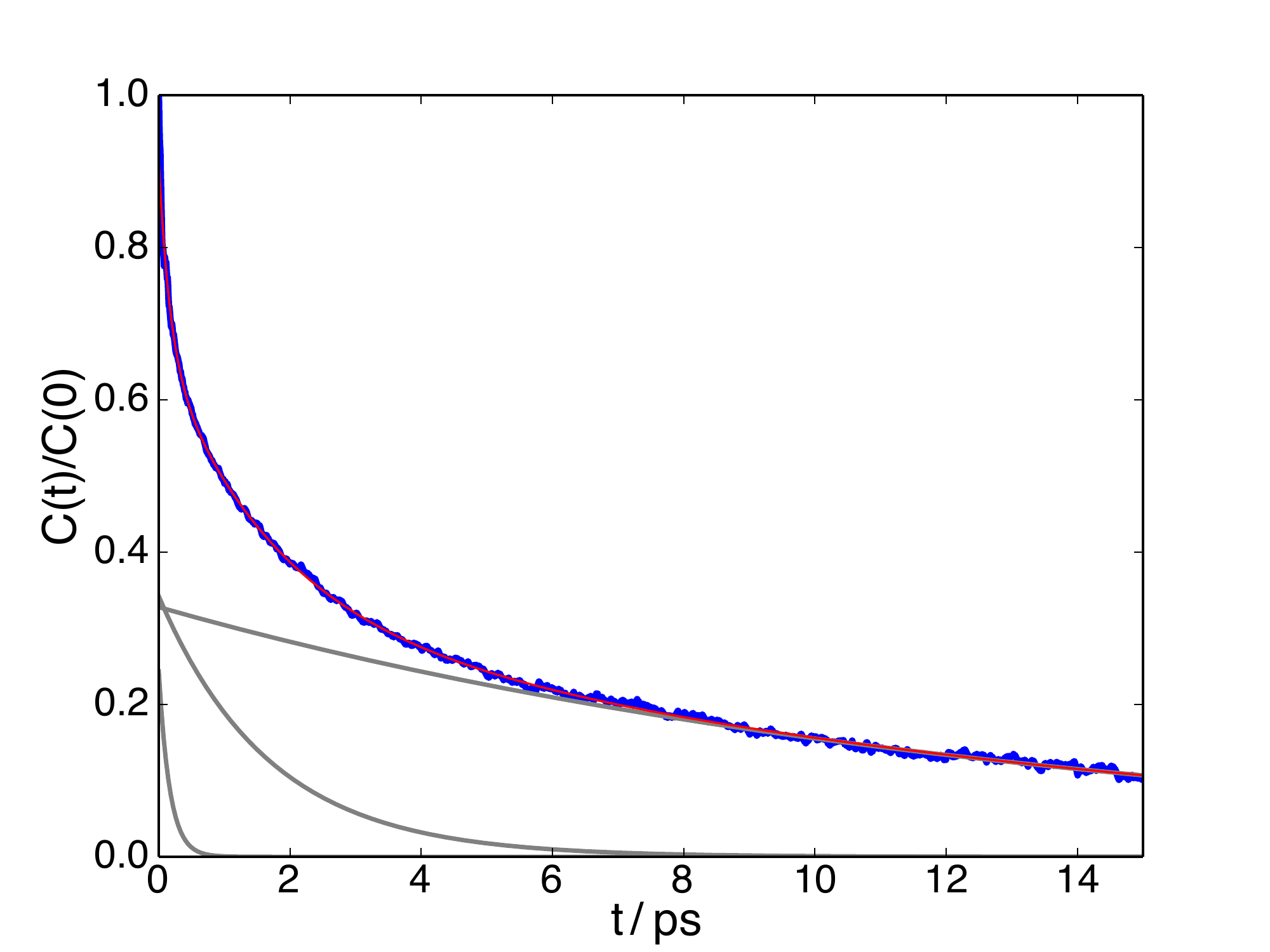}
\caption{ Rotational correlation function (blue) of the NH bond vector and the multi-exponential fit used to calculate $\tau_{\rm NH}$ (red). The individual exponential functions are shown in grey.} 
\label{fig:rotTCF}
\end{figure} 
Further insight into this conclusion comes from the MD simulations.  The rotational correlation function $C(t)$, calculated according to (\ref{eq:rotTCF}), is presented in figure~\ref{fig:rotTCF}. $C(t)$ shows an initially fast decay, followed by a much slower decay, which doesn't reach zero in the presented time interval. For the following analysis $C(t)$ has been modeled by a tri-exponential function, i.e. $C(t) = \sum_{i=1}^3 a_i \exp(- t/\tau_i )$. The resulting three time scales (amplitudes) are $\tau_1$=0.167~ps (0.24), $\tau_2$=1.69~ps (0.34) and $\tau_3$=13.4~ps (0.33). The rotational correlation time according to (\ref{eq:tau}) for this tri-exponential fit follows as 4.4 ps, i.e. it is about four times smaller than the experimental one of 16.1~ps. However, the simulations clearly confirm the order of magnitude for EAN as compared to the TEA-based systems. We note in passing, that force field MD simulations  of PILs I-III  by Strauch et al. yielded correlation times, which for II and III substantially exceeded the experimental values~\cite{strauch16_17788}.  

The three different decay times contributing to $C(t)$ can be discussed in the context of the findings by Hunger et al., who reported on dielectric relaxation and femtosecond IR experiments~\cite{hunger12_3034}.  Focussing on the transient dynamics of the initially excited ND-stretching vibration, they observed three time scales at 298~K, i.e. 0.2~ps, 2.5~ps, and 15.7~ps. The former two have been related to vibrational energy relaxation via intermediate vibrational states.  The slowest one was attributed to reorientational dynamics of the HBs.  By comparison with the time constant obtained from dielectric relaxation, they proposed a cooperative reorientation through large angle jumps around the CN bond axis \cite{hunger12_3034}. Comparison with the present correlation function analysis yields a good agreement of the slowest component of the tri-exponential fit (13.4~ps) with the above time scale (15.7~ps). However, we must note that the large angle jumps, which were considered to be at the origin for this time scale in \cite{hunger12_3034}, have not been clearly observed in the present simulations.

Relating the two shorter time scales of \cite{hunger12_3034} to the present findings is not that straightforward, although the agreement is striking. Of course, one can argue that any change in HB geometry leading to the loss of orientational correlation is connected to anharmonic couplings, which could be responsible for the observed vibrational relaxation. This is supported by the fact that the $\tau_1=0.167$~ps decay time is equal to the continuous lifetime $\tau_{\mathrm{cont}}$ of a HB in EAN. Hence, this time scale reflects the simultaneous effect of HB widening/shortening and NH-bending, which is driven by anharmonic couplings. In passing we note that the disruption of the HBs can also be viewed as being due to librational motion. 

\section{Summary}
\label{sec:summary}
We provide valuable information about the structure and dynamics of the HB network in EAN, which can be regarded as archetype of protic ionic liquids.  The properties of the HBs have been analyzed using simulations and experiments within a multi-dimensional approach, which focussed on geometric correlations, IR, and NMR spectroscopy. DFTB-based molecular dynamics provided the frame for unraveling molecular details behind the macroscopic observables. EAN features a flexible three-dimensional network of HBs with moderate strength. This makes EAN  distinct from related TEA-based PILs, which have a one-dimensional network only. Geometric correlations between HB  and NH bond lengths are not very prononced, which can be seen as a consequence of the rapid interchange of HB partners, expressed by the rather short continuous lifetime. Signatures of H-bonding are also observed in the MIR and FIR spectra as red and blue-shifted bands, respectively. DFTB has the advantage of providing a means to calculate IR spectra due to molecular fragments in a very efficient way, which allowed to unravel to origin of broad spectral bands. This way we could show that the width of NH vibrational contributions spans the range from 2800 to 3250~\cm. Rotational correlation of the NH bond vectors within the HB provides an additional means for scrutinizing the HB network. Whereas the deuteron quadrupole coupling constants for EAN and TEA-based PILs suggest comparable cation-anion interaction strengths, the dynamics is strikingly different as is manifest in the order of magnitude difference of the respective rotational correlation times. 
\section*{Acknowledgments}
The authors thank the Deutsche Forschungsgemeinschaft (DFG) for financial support through the Sfb 652. 

\section*{References}
\providecommand{\newblock}{}

\end{document}